\documentclass[aps,amsfonts,amsmath,prd,preprint,nofootinbib]{revtex4}
\newcommand{\beq}{\begin{equation}}
\newcommand{\eeq}{\end{equation}}

\usepackage{epsfig,bbm,cancel,xcolor}
\usepackage[breaklinks=true]{hyperref}

\begin{document}

\title{Classical defects in higher-dimensional Einstein gravity coupled to nonlinear $\sigma$-models}%Higher-dimensional noncanonical global defects black hole and compactification}%Some solutions around DBI global monopole and global $k$-monopole in $D$-dimensional spacetime}

\author{Ilham Prasetyo}
\email{ilham.prasetyo@sci.ui.ac.id}
\author{Handhika S. Ramadhan}
\email{hramad@ui.ac.id}
\affiliation{Departemen Fisika, FMIPA, Universitas Indonesia, Depok 16424, Indonesia.}

\def\changenote#1{\footnote{\bf #1}}

\begin{abstract}
We construct solutions of higher-dimensional Einstein gravity coupled to nonlinear $\sigma$-model with cosmological constant. The $\sigma$-model can be perceived as exterior configuration of a spontaneously-broken $SO(D-1)$ global higher-codimensional ``monopole". Here we allow the kinetic term of the $\sigma$-model to be noncanonical; in particular we specifically study a quadratic-power-law type. This is some possible higher-dimensional generalization of the Bariola-Vilenkin (BV) solutions with $k$-global monopole studied recently. The solutions can be perceived as the exterior solution of a black hole swallowing up noncanonical global defects. Even in the absence of comological constant its surrounding spacetime is asymptotically non-flat; it suffers from deficit solid angle. We discuss the corresponding horizons. For $\Lambda>0$ in $4d$ there can exist three extremal conditions (the cold, ultracold, and Nariai black holes), while in higher-than-four dimensions the extremal black hole is only Nariai. For $\Lambda<0$ we only have black hole solutions with one horizon, save for the $4d$ case where there can exist two horizons. We give constraints on the mass and the symmetry-breaking scale for the existence of all the extremal cases. In addition, we also obtain factorized solutions, whose topology is the direct product of two-dimensional spaces of constant curvature ($M_2$, $dS_2$, or $AdS_2$) with (D-2)-sphere. We study all possible factorized channels.
\end{abstract}

\maketitle
\thispagestyle{empty}
%\section{Introduction}
\setcounter{page}{1}

\section{Introduction}

General relativity coupled to $SO(3)$ global monopole was first studied by Barriola and Vilenkin~\cite{Barriola:1989hx}. They found that the exterior monopole spacetime can be brought to ``look like" Minkowski with a deficit solid angle ($\Delta=8\pi G\eta^2$), only that it is not locally flat. The size of the monopole core is of the order of $\delta\sim\eta^{-1}$. When the core is smaller than its corresponding Schwarzschild radius, their solution describes a global monopole eaten up by a black hole~\cite{Dadhich:1997mh}. The Barriola-Vilenkin (BV) monopole solution was later extended to higher-codimensional defects by Olasagasti and Vilenkin (OV)~\cite{Olasagasti:2000gx}. They studied spacetime around higher-dimensional $SO(n)$ global defects, where the hedgehog configuration wraps the $n$ extra dimensions. Upon rescaling, their solution (with $\Lambda=0$) of the spacetime transverse to the $p$-brane describes the generalization of BV metric, with deficit angle that grows with the dimension. The black hole version of this solution is described by Tangherlini metric~\cite{Tangherlini:1963bw}. These facts seem to support the well-known conjecture of ``black hole has no scalar hair" (for example, see~\cite{Coleman:1991ku}).

In field theory there has been a recent growing interest on defects solutions having noncanonical kinetic term(s), dubbed $k$-defects~\cite{Babichev:2006cy, Babichev:2007tn}. One of the simplest model of $k$-monopole is by having a power-law kinetic terms. The gravitational field of such objects and the geodesic of a test particle around it have been studied numerically in~\cite{Jin:2007fz}. Recently we have investigated analytically the corresponding black hole  and compactification solutions~\cite{Prasetyo:2015bga, Prasetyo:2016nng}. We found that the noncanonical nature of the theory allows the existence of a black hole with a genuine scalar charge, that cannot be brought into Schwarzschild upon rescaling. This solution behaves a lot like Reissner-Nordstrom, having two horizons that can be extremal or succumb to naked singularity, depending on the value of the ratio between the charge and the coupling to nonlinearity, ${\eta^4\over\beta^2}$.  

In this work we generalize our analysis to higher dimensions. We look for exact solutions of spacetime around a noncanonical $SO(D-1)$ global defects with cosmological constant in $(D+1)$ dimensions. To achieve this purpose, in Section II we consider the Einstein's equations with nonlinear $\sigma$-model as a source. In this work we limit our investigation only for the case of quadratic-power-law kinetic term. The case for Dirac-Born-Infeld (DBI) kinetic term is reported in the subsequent report. The solution with the hedgehog ansatz can be perceived as a metric around a global defects. In Section III we present the black hole solution and study its various extremal conditions. Section IV is devoted to the factorized solutions, where we analyze all possible topologies and give classical conditions of how each can be achieved. We summarize our conclusions in Section V. 
%
%_________________________________
%\clearpage
\section{Gravity of power-law global defects}

The simplest power-law global monopole theory is described by the following action
%\begin{equation}
%\mathcal{S}=\int d^{D}x \; \sqrt{|g|} \;
%\left( {R-2\Lambda \over 16\pi G} 
%-\mathcal{X}-\beta^{-2} \mathcal{X}^2 -{\lambda\over 4} \left( \vec{\Phi}^2-\eta^2 \right)
%\right), 
%\end{equation}
\begin{equation}
\mathcal{S}=\int d^{D}x \; \sqrt{|g|} \;
\left( {R-2\Lambda \over 16\pi G} 
+\mathcal{K}(\mathcal{X})-{\lambda\over 4} \left( \vec{\Phi}^2-\eta^2 \right)
\right), 
\end{equation}
with $\mathcal{X}\equiv -(1/2)\partial_M\vec{\Phi}\partial^M\vec{\Phi}$ and $\mathcal{K}$ is a functional of $\mathcal{X}$ only. Here $G$ and $\Lambda$ are the $D$-dimensional Newton's constant and cosmological constant, repsectively. As is discussed in~\cite{Babichev:2006cy, Jin:2007fz}, the form of $\mathcal{K}(\mathcal{X})$ must satisfy the following:
\begin{eqnarray}
\mathcal{K}(\mathcal{X})=\begin{cases}
-\mathcal{X},\ \ \  |\mathcal{X}|\ll1,\\
-\mathcal{X}^{\alpha},\ \ \ |\mathcal{X}|\gg1,
\end{cases}
\end{eqnarray}
where $\alpha$ is some constant. The upper condition ensures that the theory goes to its canonical counterpart in its perturbative regime to avoid ``zero-kinetic problem". In this work, we particularly choose the following specific form:
\begin{equation}
\mathcal{K}(\mathcal{X})\equiv-\mathcal{X}-\beta^{-2} \mathcal{X}^2,
\label{eq:kbeta}
\end{equation}
with $\beta$ is the parameter coupling to nonlinearity. In the limit of $\beta\rightarrow\infty$ the theory reduces to the canonical Einstein-$\sigma$-model case. This model can be perceived as the simplest nonlinear $\sigma$-model with noncanonical kinetic term(s), and can be thought of as a truncation of a more general nonlinear lagrangian. Note that this truncation is only valid as long as $|\mathcal{X}|\ll\beta^{-1}$. Otherwise, higher-order correction terms should be included. We surely do not expect this truncated toy model to represent a realistic theory. However, we can regard this as a model that is simple enough to study while at the same time gives, as we shall see, genuine features of solutions not present in its canonical counterpart. Another type of toy model we can  consider is the Dirac-Born-Infeld (DBI)\footnote{Note that the lagrangian~\eqref{eq:kbeta} cannot be obtained by expanding lagrangian~\eqref{eq:kdbi} and truncate it up to second order, since the sign in front of the second order is different, $b^2\left(1-\sqrt{1+{2\mathcal{X}\over b^2}}\right)=-\mathcal{X}+{1\over2b^2}\mathcal{X}^2+O(\mathcal{X}^3)$. Thus we may perceive the toy model~\eqref{eq:kbeta} as distinct from \eqref{eq:kdbi}.} global monopole coupled to gravity,
\begin{equation}
\mathcal{K}(\mathcal{X})\equiv b^2\left(1-\sqrt{1+{2\mathcal{X}\over b^2}}\right).
\label{eq:kdbi}
\end{equation}
Note that the corresponding effective action can be written as
\begin{equation}
\mathcal{S}=\int d^{D}x~ \sqrt{|g|} \left[
{R-2\lambda\over 16\pi G}-
\beta^2\sqrt{1-{\partial_M \vec{\Phi} \partial^M \vec{\Phi} \over \beta^2}}
-{\lambda\over4}\left(
\vec{\Phi}^2-\eta^2\right)^2\right],
\end{equation}
with $\lambda\equiv\Lambda-16\pi G\beta^2$ is the effective cosmological constant. This type of theory is at the moment being studied and shall be reported separately~\cite{nextpub}.

The scalar field part of the action admits an $SO(d)$ symmetry which is spontaneously broken to $SO(d-1)$, where $d$ is the number of degrees of freedom of the scalar field $\vec{\Phi}$. The nonlinear $\sigma$-model constraint restricts the scalar field to stay in its vacuum manifold $\mathcal{M}$, defined by $\vec{\Phi}^2=\eta^2$, which is an $S^{d-1}$. The scalar field can then be regarded as having internal coordinates, $\vec{\Phi}=\vec{\Phi}\left(\phi^i\right)$, $i=1, 2,\cdots, d-1$. Its effective action is then given by
\begin{equation}
\mathcal{S}=\int d^{D}x \; \sqrt{|g|} \; 
\left[ {R-2\Lambda \over 16\pi G} 
-X-\beta^{-2} X^2
\right] ,
\end{equation}
with $X\equiv-(1/2)\eta^2 h_{ij}\partial_M{\phi}^i\partial^M{\phi}^j$, where $h_{ij}=h_{ij} (\phi^k)\equiv{\partial\vec{\Phi}\over\partial\phi^i}\cdot{\partial\vec{\Phi}\over\partial\phi^j}$ is the internal metric on the manifold $\mathcal{M}$.

The equation of motion for the scalar field and the energy-momentum tensor are, respectively, 
\begin{equation}
\label{eq:sigmamod}
{1\over\sqrt{|g|}}\partial_M\left(
\sqrt{|g|}(1+2\beta^{-2}X)\eta^2 h_{ij} \partial^M \phi^j
\right)=(1+2\beta^{-2}X){\eta^2\over2} \partial_M\phi^m\partial^M\phi^n {\partial h_{mn}\over\partial\phi^i},
\end{equation}
\begin{equation}
T^M_N=\delta^M_N \left[ {\Lambda\over 8\pi G} 
+X+\beta^{-2} X^2 \right]+(1+2\beta^{-2}X) \eta^2 h_{ij} \partial^M\phi^i \partial_N\phi^j.
\end{equation}
They are to be solved along with the Einstein's equations.

In this work we seek solutions for spherically-symmetric metric
\begin{equation}
ds^2=A^2(r)dt^2 -B^2(r)dr^2-C^2(r)d\Omega^2_{D-2}.\label{eq:metric}
\end{equation}
The unit $(D-2)$-sphere are parametrized by the angular coordinates $\theta^1,\theta^2,...,\theta^{D-2}$. The Ricci tensor components are
\begin{eqnarray}
R^t_t&=&B^{-2}\left[
{A''\over A}-{A'B'\over AB} +(D-2) {A'C'\over AC}
\right],\\
R^r_r&=&B^{-2}\left[
{A''\over A} + (D-2) {C''\over C} -{B'\over B}\left\{
{A'\over A} +(D-2) {C'\over C}
\right\}\right],\\
R^\theta_\theta&=&B^{-2}\left[
{C''\over C}+{C'\over C}\left\{
{A'\over A}-{B'\over B}+(D-3){C'\over C} \right\}\right]-{(D-3)\over C^2}.
\end{eqnarray}
For the sake of later purpose, we also show the components of Einstein tensor
\begin{eqnarray}
G^0_0&=&-{(D-2)\over B^2}{C''\over C}+{(D-2)\over B^2}{B'C'\over BC}+{(D-2)(D-3)\over2C^2}\left(1-{C'^2\over B^2}\right),\\
G^r_r&=&-{(D-2)\over B^2}{A'C'\over AC}+{(D-2)(D-3)\over2C^2}\left(1-{C'^2\over B^2}\right),\\
G^\theta_\theta&=&-{1\over B^2}{A''\over A}-{(D-3)\over B^2}{C''\over C}+{1\over B^2}{A'B'\over AB}-{(D-3)\over B^2}{A'C'\over AC}+{(D-3)\over B^2}{B'C'\over BC}\nonumber\\
&&+{(D-4)(D-3)\over2C^2}\left(1-{C'^2\over B^2}\right).
\end{eqnarray}
The crucial part in this work comes when choosing the appropriate ansatz for the scalar field. Here we follow~\cite{GellMann:1984sj} in taking the simplest ansatz that respects spherical symmetry,
\begin{equation}
\phi^i(\theta^i)=\theta^i,
\end{equation}
where now $i=1, \cdots, D-2$; the number of degrees of freedom of the internal space should equal the angular degrees of freedom of the coordinate space. It can be seen that such an ansatz satisfies Eq.\eqref{eq:sigmamod} trivially if
\begin{equation}
h_{ij}=-{1\over C^2}g_{ij}.
\end{equation}
This will give us
\begin{equation}
T^0_0=T^r_r=\left[ {\Lambda\over 8\pi G} 
+X+\beta^{-2} X^2 \right],
\end{equation}
\begin{equation}
T^\theta_\theta=T^0_0-(1+2\beta^{-2}X)  {\eta^2\over C^2},
\end{equation}
with $X\equiv(D-2)\eta^2/2C^2$. In the next sections we shall discuss several classes of solutions to these equations.

\section{Black hole solutions}

Let us first consider an ansatz:
\begin{equation}
C(r)\equiv r.
\end{equation}
The Einstein's equations, $R^A_{B}=8\pi G\left(T^A_B-\delta^A_B {T\over (D-2)}\right)$, give us
\begin{eqnarray}
&& R^0_0={1\over B^2}\left[ {A''\over A} - {A'B'\over AB} + (D-2) {A'\over rA}
\right]=-{2\over D-2}\Lambda + {(D-2)\over 2} {8\pi G\eta^4\over \beta^2 r^4},\\
&& R^r_r={1\over B^2}\left[ {A''\over A} - {A'B'\over AB} - (D-2) {B'\over rB}
\right]=-{2\over D-2}\Lambda + {(D-2)\over 2} {8\pi G\eta^4\over \beta^2 r^4},\\
&& R^\theta_\theta = {1\over B^2}\left[ {A'\over rA} - {B'\over rB} + {D-3\over r^2}
\right] -{D-3\over r^2}\label{eq:ttheta} 
=-{2\over D-2}\Lambda - {8\pi G\eta^2\over r^2} - {(D-2)\over 2} {8\pi G\eta^4\over \beta^2 r^4}.
\end{eqnarray}
From $R^0_0-R^r_r$ we have
\[{A'\over A}+{B'\over B}=0 \Rightarrow (AB)'=0.\]
Without loss of generality we can take
\begin{equation}
A=B^{-1}.
\end{equation}
Substituting into $R^\theta_\theta$ yields
\[R^\theta_\theta = {1\over r^{D-2}}\left( {r^{D-3}\over B^2} \right)'- {D-3\over r^2},\]
and substituting it back to~\eqref{eq:ttheta}, we obtain the following solution
\begin{eqnarray}
B^{-2}=1-{8\pi G\eta^2\over(D-3)}-{2\Lambda r^2\over (D-2)(D-1)} 
-{4(D-2)\pi G\eta^4\over (D-5)\beta^2 r^2}-{2GM\over r^{(D-3)}},
\end{eqnarray}
with $M$ a constant of integration. As in the case of BV and OV solutions, ours is only valid for $D>3$. The noncanonical nature of the scalar field, $X^2$, adds another constraint that $D\neq 5$. 

To better see what this solution tells, let us rescale\footnote{In general, an arbitrary metric solution $ds^2=A(r) dt^2-{dr^2\over A(r)}-r^2d\Omega^2_{D-2}$ in the form of $A(r)=1-\Delta-\sum^{\infty}_{i=1}\alpha_i r^i-\sum^{\infty}_{j=1}\sigma_j r^{-j}$ where $\alpha_i$ and $\sigma_j$ are constants can be brought to $ds^2=A(r) dt^2-{dr^2\over A(r)}-\left(1-\Delta\right)r^2d\Omega^2_{D-2}$ with $A(r)\rightarrow 1-\sum^{\infty}_{i=1}\alpha_i r^i-\sum^{\infty}_{j=1}\sigma_j r^{-j}$, should we simultaneously also transform $\alpha_i\rightarrow\alpha_i\left(1-\Delta\right)^{i-2\over 2}$ and $\sigma_j\rightarrow\sigma_j\left(1-\Delta\right)^{-(j+2)\over 2}$. See~\cite{Marunovic:2013eka}.} $t\rightarrow t(1-\Delta)^{1/2}$ and $r\rightarrow r(1-\Delta)^{-1/2}$, where $\Delta\equiv 8\pi G\eta^2/(D-3)$ is the deficit solid angle. The metric thus becomes
\begin{equation}
ds^2=f(r)dt^2-{dr^2\over f(r)}-(1-\Delta)r^2d\Omega^2_{(D-2)},\label{eq:metricsol}
\end{equation}
with
\begin{eqnarray}
f(r)=1-{2\Lambda r^2\over (D-2)(D-1)} 
-{4(D-2)\pi G\eta^4\over (D-5)\beta^2 r^2}-{2GM\over r^{(D-3)}},\label{eq:solblackhole}
\end{eqnarray}
where simultaneously we also transform $M\rightarrow M(1-\Delta)^{(1-D)/2}$ and $\beta\rightarrow\beta(1-\Delta)$. The Kretchmann scalar $K^2\equiv R_{ABCD} R^{ABCD}$ yields
\begin{eqnarray}
K^2&=&\frac{4 (D-2)}{B^4}\left[{\left(\frac{A'}{Ar}\right)^2}+{\left(\frac{B'}{Br}\right)^2}\right]+\frac{4 }{B^4}{\left(\frac{A''}{A}-\frac{A'B'}{AB}\right)^2}
+\frac{8 (D-2) (D-3)}{r^4}\left(1-\frac{1}{B^2}\right)^2 \nonumber \\
&=&{\left[{2  (D-2)^2 \left(D^3-9 D^2+23 D-15\right) G M \over { (D-5) (D-2) (D-1)} r^{D-1}}+{ 24 (D-2)^2 (D-1) \eta ^4 \pi G +4 \beta ^2 (D-5) \Lambda  r^4\over {\beta ^2 (D-5) (D-2) (D-1)} r^{4}}\right]^2}\nonumber\\
&&+\frac{8 (D-3) (D-2)}{r^4} \left(\frac{2 \Lambda  r^2}{(D-2)(D-1)}+\frac{8\pi G\eta ^2 }{D-3}+{2 G M \over r^{D-3}}+\frac{8(D-2) \eta ^4\pi G }{2 \beta ^2 (D-5) r^2}\right)^2\nonumber\\
&&+\frac{2 (D-2)}{r^2} \left(-\frac{4 \Lambda  r}{(D-2)(D-1)}+{2 (D-3) G M \over r^{D-2}}+\frac{8(D-2) \eta ^4\pi G }{\beta ^2 (D-5) r^3}\right)^2,
\end{eqnarray}
which implies absolute singularity at $r=0$.

The metric~\eqref{eq:metricsol} describes a ``scalarly-charged" (A)dS-Tangherlini black hole with global monopole. The charge is genuinely due to the nonlinearity of the scalar field's kinetic term. In the limit of $\beta\rightarrow\infty$ Eq~\eqref{eq:solblackhole} reduces to the ordinary (A)dS-Tangherlini black hole and the scalar charge disappears. The difference between this solution and (A)dS-Reissner-Nordstrom-Tangherlini solution is that the latter  has the EM charge term that goes like $O(r^{-2(D-3)})$ while in the former the scalar charge term has a fixed polynomial order, $O(r^{-2})$, regardless of the dimensions. It is amusing that this scalar-charge term is very similar to the magnetic term in the metric solution of the generalized Nariai black hole with multiple magnetic charges~\cite{Batista:2016qnu, Ortaggio:2007hs}. For $D=4$ the charge term is positive, much like the case with electric (or magnetic) charge in Reissner-Nordstrom solution. This looks like a solution that violates the well-known black-hole-has-no-hair theorem, but the reader should be reminded that the spacetime around this blackhole is not asymptotically flat due to the deficit angle; it has conical solid angle. Such a metric solution has been discovered and studied (albeit without the noncanonical kinetic term), for example, in~\cite{Dadhich:1997mh, Jensen:1995fz, Yu:1994fy, Lustosa:2015hwa, Mazharimousavi:2014uya}\footnote{Black holes solutions with other type of topological defects ({\it e.g.,} cosmic strings) can be found in~\cite{Aryal:1986sz, Achucarro:1995nu}. There, the solutions exhibit similar property of conical deficit angle.}. The noncanonical nature of this solution allows it to evade the no-scalar-hair theorem. The static black hole solution ceases to exist when 
\begin{equation}
\eta>\eta_{crit}\equiv\sqrt{D-3\over8\pi G}.\label{eq:etacrit}
\end{equation}
It should be pointed out that the result~\eqref{eq:etacrit} above is generic to the theory of global monopole; It is the property of the corresponding vacuum manifold (the infra-red regime), independent of the noncanonicality of the kinetic term (the UV-regime). It is, however, interesting to stress that such deficit angle only appears in the black hole having spherical topology. One of us (HSR) investigated the corresponding global monopole black hole in hyperbolic topology, and one of the peculiar properties we found is that such a black hole possesses surplus (instead of deficit) solid angle~\cite{ramadhanpradhana}. This result shall be reported elsewhere. 

We study the existence of horizons by plotting this metric solution for different coupling values, cosmological constants, and dimensions. There are several interesting cases.

\subsection{The Case with $\Lambda=0$}
 
\begin{figure}
	\centering
	\includegraphics[width=0.7\linewidth]{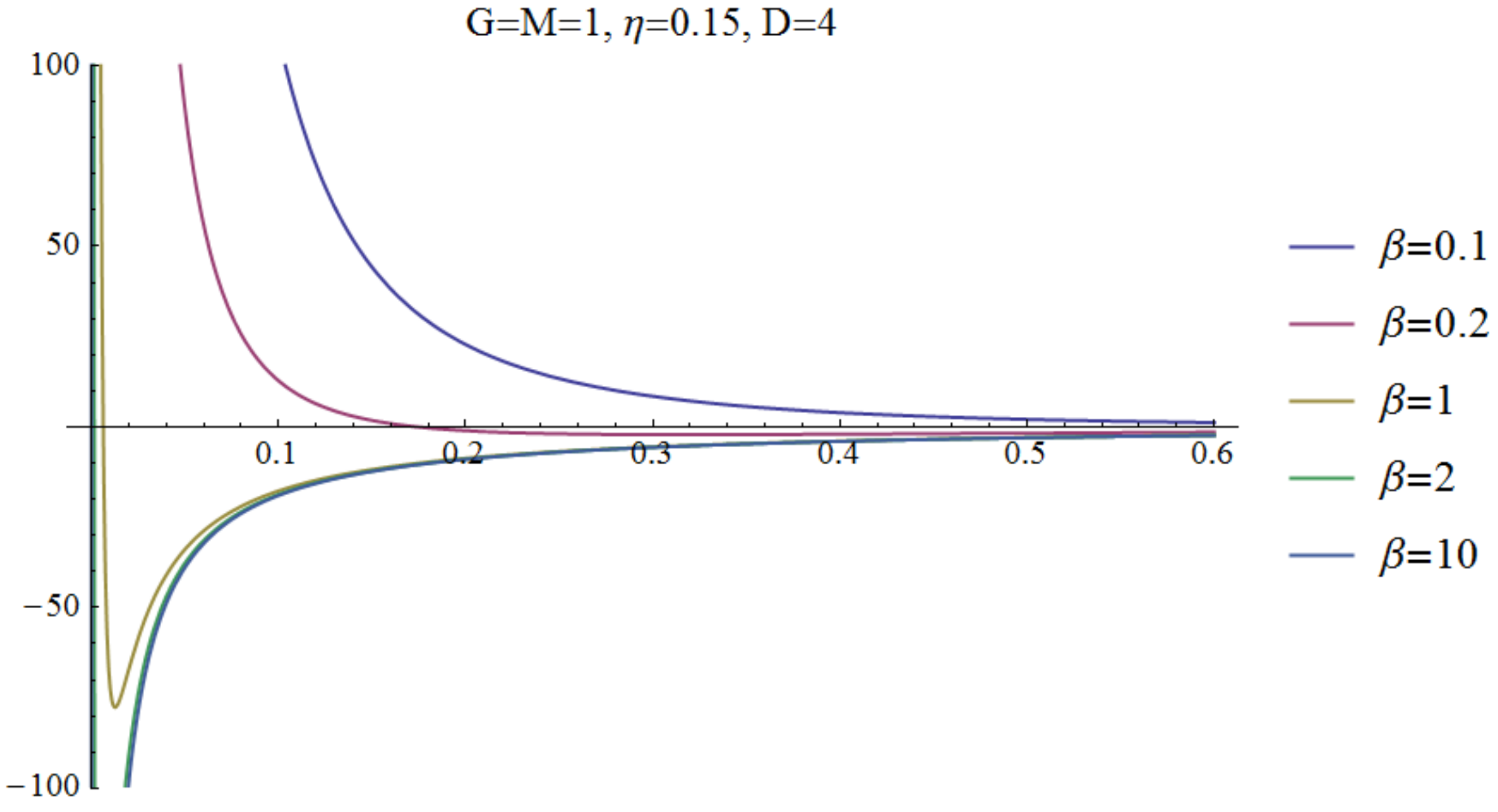}
	\caption{$f(r)$ in $D=4$ for various $\beta$. As $\beta$ increases the local minimum gets deeper. In the limit of $\beta\rightarrow\infty$ the local minimum disappears. }
	\label{fig:k-betatambah}
\end{figure}

In this limit, Eq.~\eqref{eq:solblackhole} becomes
\begin{equation}
f(r)=1-{4(D-2)\pi G\eta^4\over (D-5)\beta^2 r^2}-{2GM\over r^{(D-3)}}.\label{eq:solblackholelambdanol}
\end{equation}
The case for $D=4$ has been discussed extensively in~\cite{Prasetyo:2015bga, Prasetyo:2016nng}. This is a Reissner-Nordstrom-like black hole with two horizons
\begin{equation}
r_{\pm}=GM\left(1\pm\sqrt{1-{8\pi\eta^4\over M^2G\beta^2}}\right).
\end{equation}
\begin{figure}
	\centering
	\includegraphics[width=0.7\linewidth]{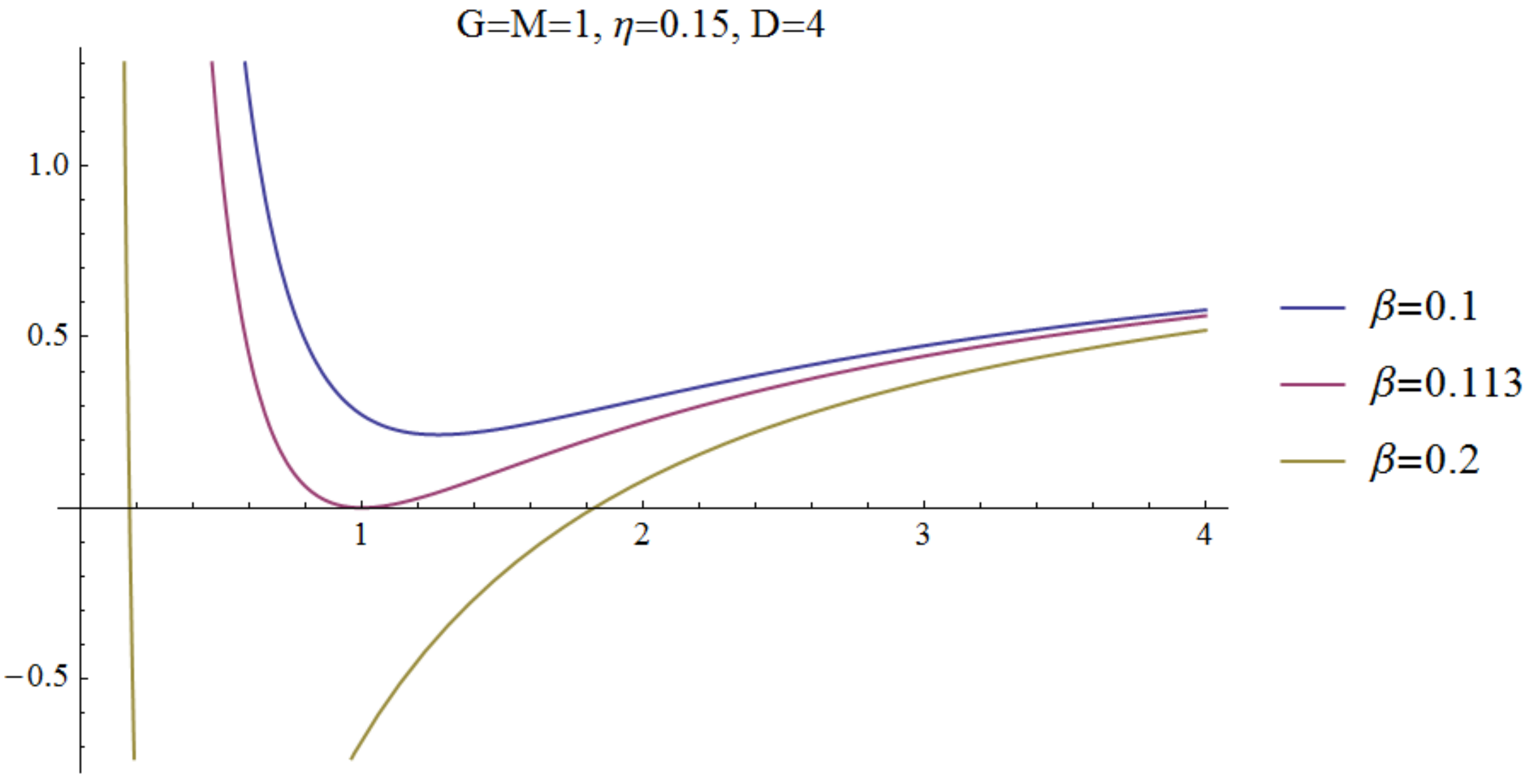}
	\caption{The four-dimensional metric $f(r)$ as a function of $r$ for the case of two horizons, extremal, and naked singularity.}
	\label{k-beta}
\end{figure}
Their behavior is shown in Figs.~\ref{k-beta} and \ref{fig:k-betatambah}.

\begin{figure}
	\centering
	\includegraphics[width=0.7\linewidth]{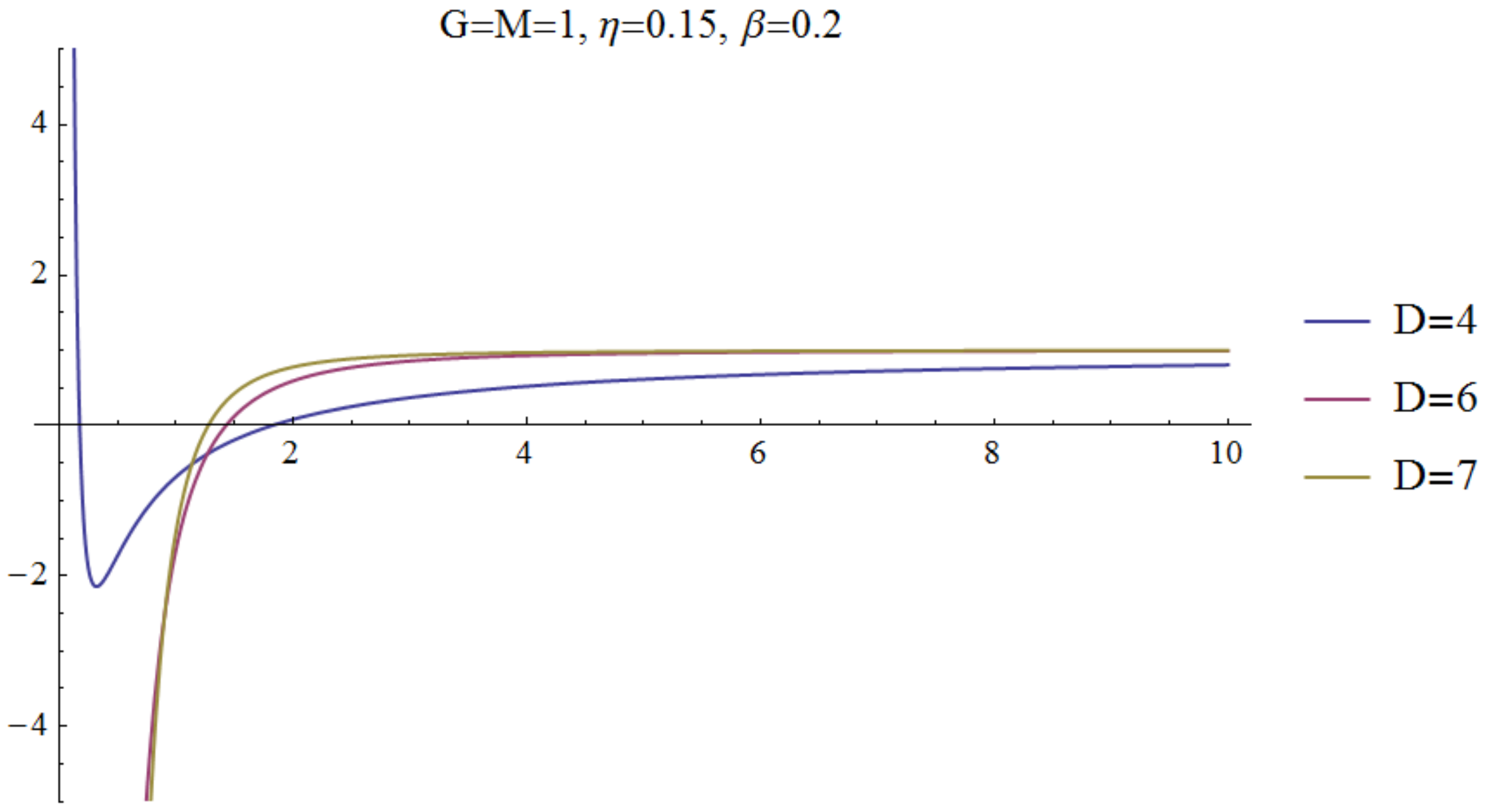} (c)
	\caption{The D-dimensional metric $f(r)$ for several $D$.}
	\label{fig:k-beta-lambda-0}
\end{figure}
For $D>5$, it is in general difficult to obtain the exact roots of $f_D(r)=0$. From Eq.~\eqref{eq:solblackholelambdanol} we can see that since the second term is always negative there is no local minimum of $f_D(r)$. Thus there is only one real root for each $D$; the higher-dimensional black hole with noncanonical global monopole all resemble Tangherlini's. This can be seen in Fig.~\ref{fig:k-beta-lambda-0}. For example, in $D=6$ we have
\begin{equation}
f(r)=1-{2GM\over r^3}-{16\pi G\eta^4\over\beta^2 r^2}.
\end{equation}
Out of three, only one root is real, given by
\begin{equation}
r_{H}={1\over3}\bigg[a+a^2\left(2\over 2a^3+27 b+3\sqrt{3}\sqrt{4a^3b+27b^2}\right)^{1/3}+\left(2a^3+27b+3\sqrt{3}\sqrt{4a^3b+27b^2}\over2\right)^{1/3}\bigg],
\end{equation}
where $a\equiv{16\pi G\eta^4\over\beta^2}$ and $b\equiv2GM$. In $D=7$, we have
\begin{equation}
f(r)=1-{2GM\over r^4}-{10\pi G\eta^4\over\beta^2 r^2}.
\end{equation}
The positive roots are
\begin{equation}
r_{\pm}=\sqrt{{5G\pi\eta^4\over\beta^2}\pm{\sqrt{2GM\beta^4+25G^2\pi^2\eta^8}\over\beta^2}}.
\end{equation}
But it can easily be seen that for every physical values of $\beta$ and $\eta$, only $r_+$ is real.

\subsection{The Case with $\Lambda\neq0$}

The existence of the number of roots of~\eqref{eq:solblackhole} depends on the sign of $\Lambda$. For $D=4$, we have the polynomial equation
\begin{equation}
f(r)\equiv1-{\Lambda r^2\over3} 
+{8\pi G\eta^4\over\beta^2 r^2}-{2GM\over r}=0.
\end{equation}
This is a fourth-order polynomial. In general one expects to have four roots, though not all (or even none) of them are real. In~\cite{Prasetyo:2015bga} we discussed the corresponding horizons under the almost-purely de Sitter condition ($M=0$). One of our main results in this paper is the exact solutions of the black hole horizons without taking any approximation; that is, by setting $M\neq0$. 

For the case of $\Lambda>0$, the situation resembles the dS-Reissner-Nordstrom black hole. We can, therefore, borrow the language of~\cite{Romans:1991nq}. There exist at most three horizons: the inner ($r_-$), the outer ($r_+$), and the cosmological ($r_c$) horizons (satifsying $r_-<r_+<r_c$). Besides the non-extremal case where the black holes have all three horizons, we can also have solutions possessing some extremal conditions. They can be characterized as: $r_-=r_+$ (the cold black hole), $r_+=r_c$ (the Nariai black hole), and $r_-=r_+=r_c$ (the ultracold black hole)~\cite{Cardoso:2004uz}. 

The cold black hole can be characterized in such a way that the metric function can be written as~\cite{Romans:1991nq}
\begin{equation}
f_{cold}(r)=\left(1-{r_0\over r}\right)^2\left[1-{\Lambda\over3}(r^2+2r_0 r+3r_0^2)\right],\label{eq:extremecold}
\end{equation}
with $r_0\equiv r_-=r_+$ the common root. In this critical extremal condition, $M$, $\eta$, $\beta$, and $\Lambda$ are related through:
\begin{eqnarray}
M&=&{r_0\over G}\left(1-{2\over3}\Lambda r_0^2\right),\nonumber\\
{\eta^4\over\beta^2}&=&{r_0^2\over8\pi G}\left(1-\Lambda r_0^2\right).\label{eq:condextrem}
\end{eqnarray} 
\begin{figure}
	\centering
	\includegraphics[width=0.7\linewidth]{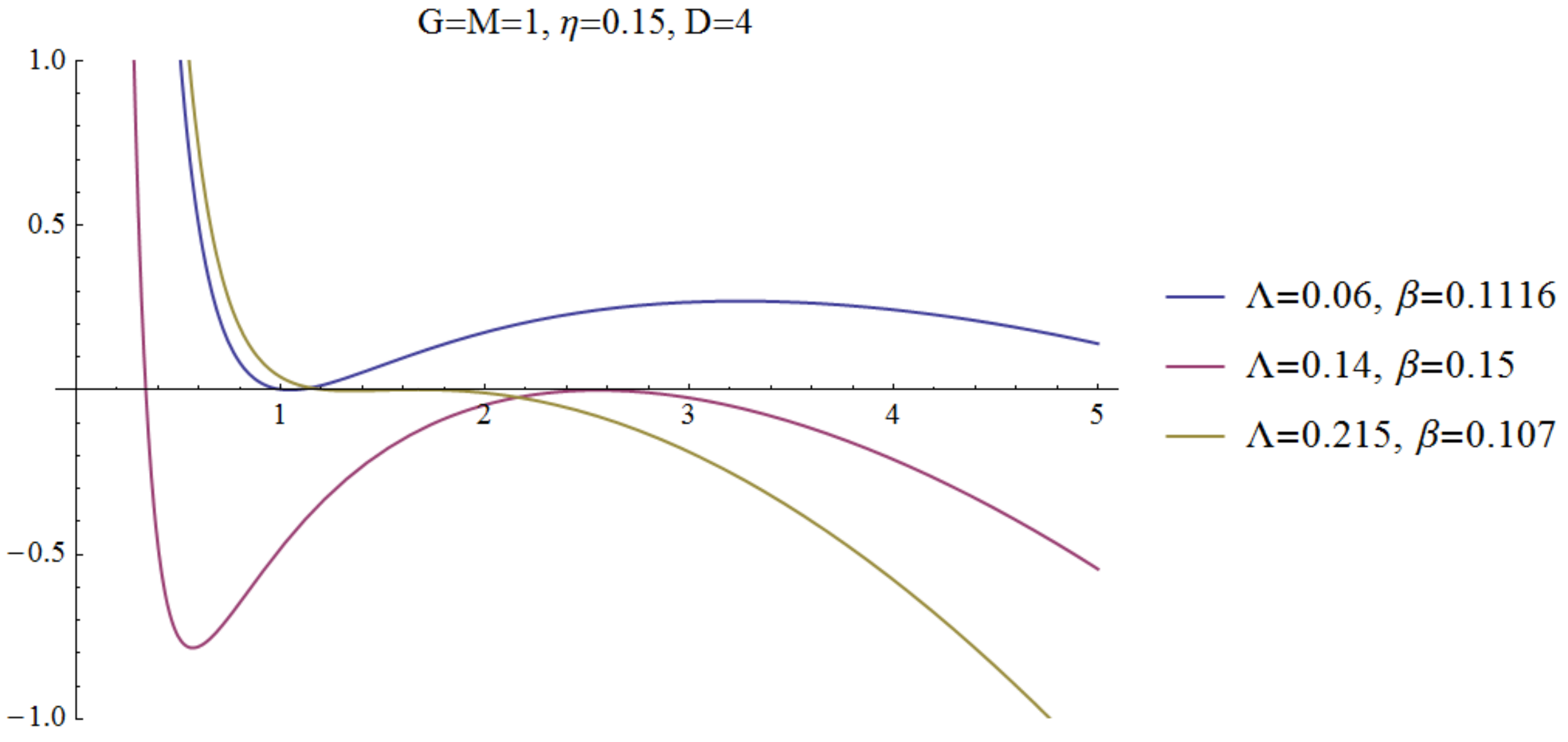}
	\caption{Transition from cold black hole ($\Lambda=0.06$ and $\beta=0.1116$) to ultracold black hole ($\Lambda=0.215$ and $\beta=0.17$) and eventually to Nariai black hole ($\Lambda=0.14$ and $\beta=0.15$).}
	\label{fig:k-cold-ultra-nariai}
\end{figure}
In order that $\eta$ and $\beta$ have physically-reasonable values we must have $r_0\leq{1\over\sqrt{\Lambda}}$. The cold black hole exists whenever $0<r_0<{1\over\sqrt{2\Lambda}}$. At $r_0={1\over\sqrt{2\Lambda}}$ we have all three horizons coincide. This is the ultracold black hole, characterized by
\begin{equation}
f_{ultracold}(r)=-{r^2\over6 r_0}\left(1-{r_0\over r}\right)^3\left(1+{3r_0\over r}\right).
\end{equation}
The mass ($M$) and the ratio of $\eta^2$ and $\beta$ are related to $\Lambda$ by
\begin{eqnarray}
M&=&{2\over\sqrt{18\Lambda}G},\nonumber\\
{\eta^2\over\beta}&=&{1\over32\pi G\Lambda}.
\end{eqnarray} 
As long as ${\beta\over\Lambda}\lesssim4$ the ultracold black hole can form before the entire solid angle is eaten up. In the range of $1/\sqrt{2\Lambda}<r_0\leq1/\sqrt{\Lambda}$ the ultracold common horizon disintegrates into $r_-$ and $r_0\equiv r_+=r_c$, and the black hole interpolates into the so-called Nariai regime~\cite{Nariai} (see Fig.~\ref{fig:k-cold-ultra-nariai}). It becomes maximal (chargeless or neutral Nariai) when $r_0={1\over\sqrt{\Lambda}}$. Here, the inner horizon $r_-$ disappears and the metric can be written as
\begin{equation}
f_{Nariai}(r)=-{1\over3r_0^2}\left(1-{r_0\over r}\right)^2\left(r^2+2r_0r\right).
\end{equation}
In this case, 
\begin{eqnarray}
M&=&{1\over3\sqrt{\Lambda}G},\nonumber\\
{\eta^2\over\beta}&=&0.
\end{eqnarray}
\begin{figure}
	\centering
	\includegraphics[width=0.7\linewidth]{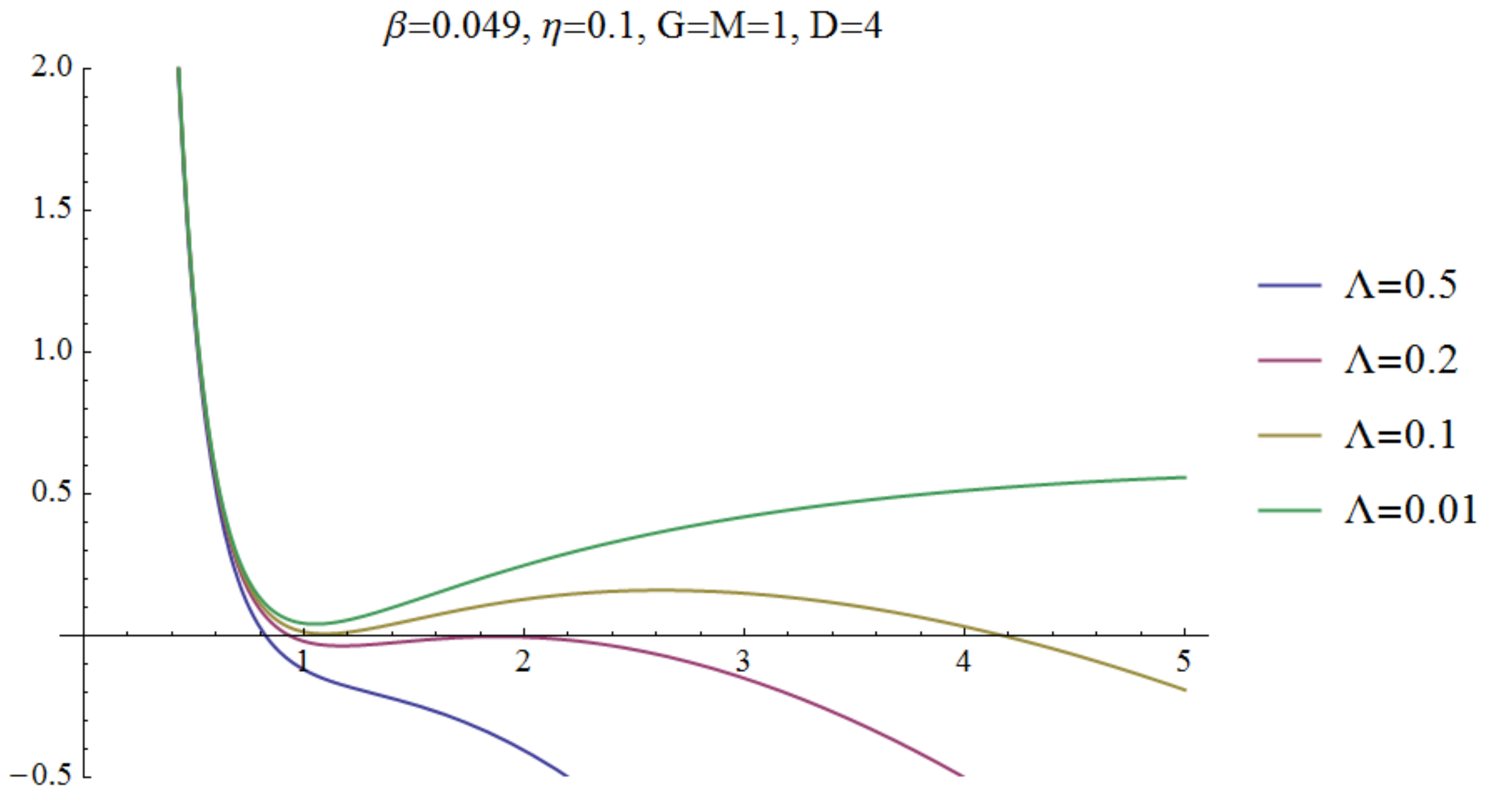}
	\caption{Black hole transition. By varying $\Lambda$ it can be seen that we can have transition from solution with one (inner) horizon, to Nariai, to cold, and finally to naked singularity.}
	\label{fig:k-transition-to-singular}
\end{figure}
In Fig.~\ref{fig:k-transition-to-singular} we can see how the black hole smoothly transitions from having one horizon to naked singularity through cold and Nariai states. 

\begin{figure}
	\centering
	\includegraphics[width=0.7\linewidth]{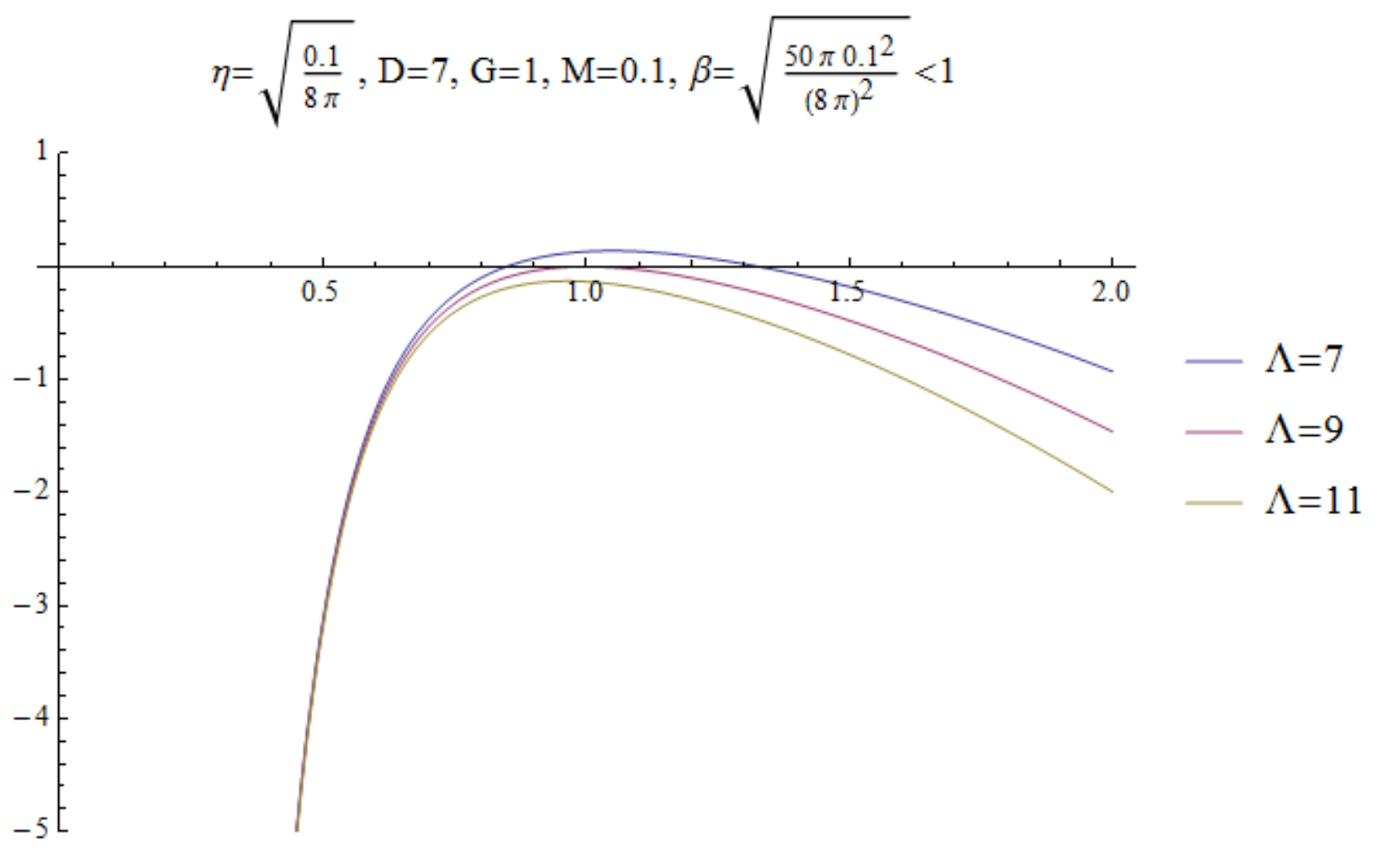}
	\caption{Transition of $7d$ black hole horizon from nonextremal, to extremal, and to naked singularity.}
	\label{fig:k-trans-d7}
\end{figure}

For $D>5$, the third term in Eq.~\eqref{eq:solblackhole} is always negative and does have a fixed power, $r^{-2}$,
\begin{equation}
f_D(r)=1-{2\Lambda r^2\over (D-2)(D-1)} 
-{4(D-2)\pi G\eta^4\over (D-5)\beta^2 r^2}-{2GM\over r^{(D-3)}}.
\end{equation}
As a result, there exist at most two horizons only. The extremal case happens when~\cite{Cardoso:2004uz}: 
\begin{equation}
f(r)=\left(1-{r_0\over r}\right)^2\left[1-{\Lambda\over3}\left(r^2+a+b r+{c_1\over r}+{c_2\over r^2}+\cdots+{c_{D-5}\over r^{D-5}}\right)\right],
\end{equation}
where determining the values of the constants $a, b, c_1,\cdots, c_{D-5}$ by matching the above equation with Eq.~\eqref{eq:solblackhole} leads to the following conditions to be satisfied
\begin{figure}
	\centering
	\includegraphics[width=0.7\linewidth]{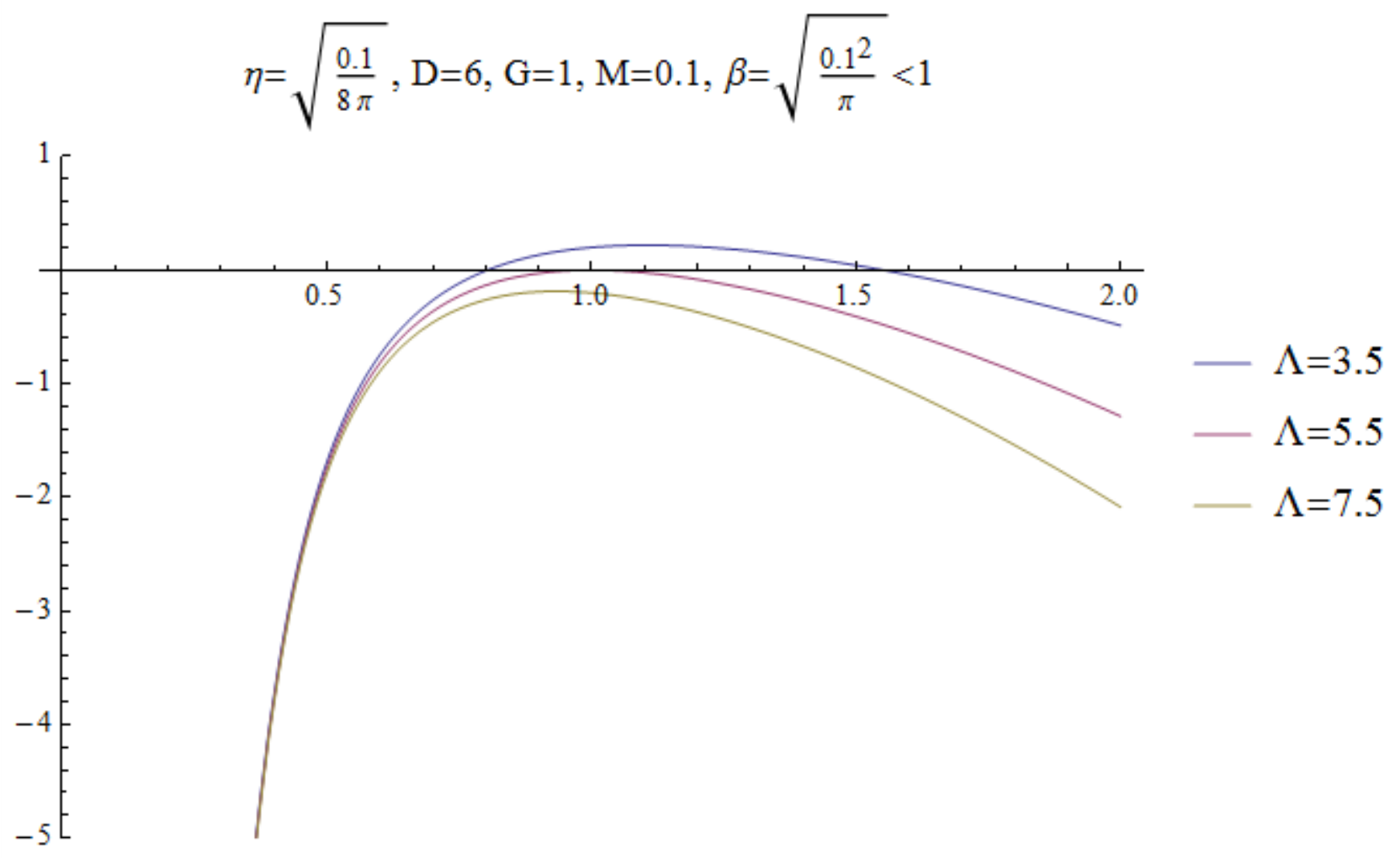}
	\caption{Transition of $6d$ black hole horizon from nonextremal, to extremal, and to naked singularity.}
	\label{fig:k-trans-d6}
\end{figure}
\begin{eqnarray}
M&=&{r_0^{D-3}\over(D-5) G}\left({4\Lambda r_0^2\over(D-1)(D-2)}-1\right),\nonumber\\
{\eta^4\over\beta^2}&=&{r_0^2\over(D-2)4\pi G}\left((D-3)-{2\Lambda\over(D-2)}r_0^2\right).\label{eq:condextremD}
\end{eqnarray}
One can easily see that putting $D=4$ reduces condition~\eqref{eq:condextremD} into~\eqref{eq:condextrem}. The positivity of mass $M$ and ${\eta^4\over\beta^2}$ ratio implies $\sqrt{(D-1)(D-2)\over4\Lambda}<r_0<\sqrt{(D-3)(D-2)\over2\Lambda}$. In Figs.~\ref{fig:k-trans-d6} and~\ref{fig:k-trans-d7} we show how transition from nonextremal black hole can happen by varying $\Lambda$ to extremal and to naked singularity in six and seven dimensions.

For $\Lambda<0$, we have\footnote{It is well-known that in $D=3$ with $\Lambda<0$ there exists a $(2+1)$-dimensional black hole solution, called BTZ black hole~\cite{Banados:1992wn, Banados:1992gq}.}
\begin{equation}
f_D(r)=1+{2|\Lambda| r^2\over (D-2)(D-1)} 
-{4(D-2)\pi G\eta^4\over (D-5)\beta^2 r^2}-{2GM\over r^{(D-3)}}.
\end{equation}
\begin{figure}
	\centering
	\includegraphics[width=0.7\linewidth]{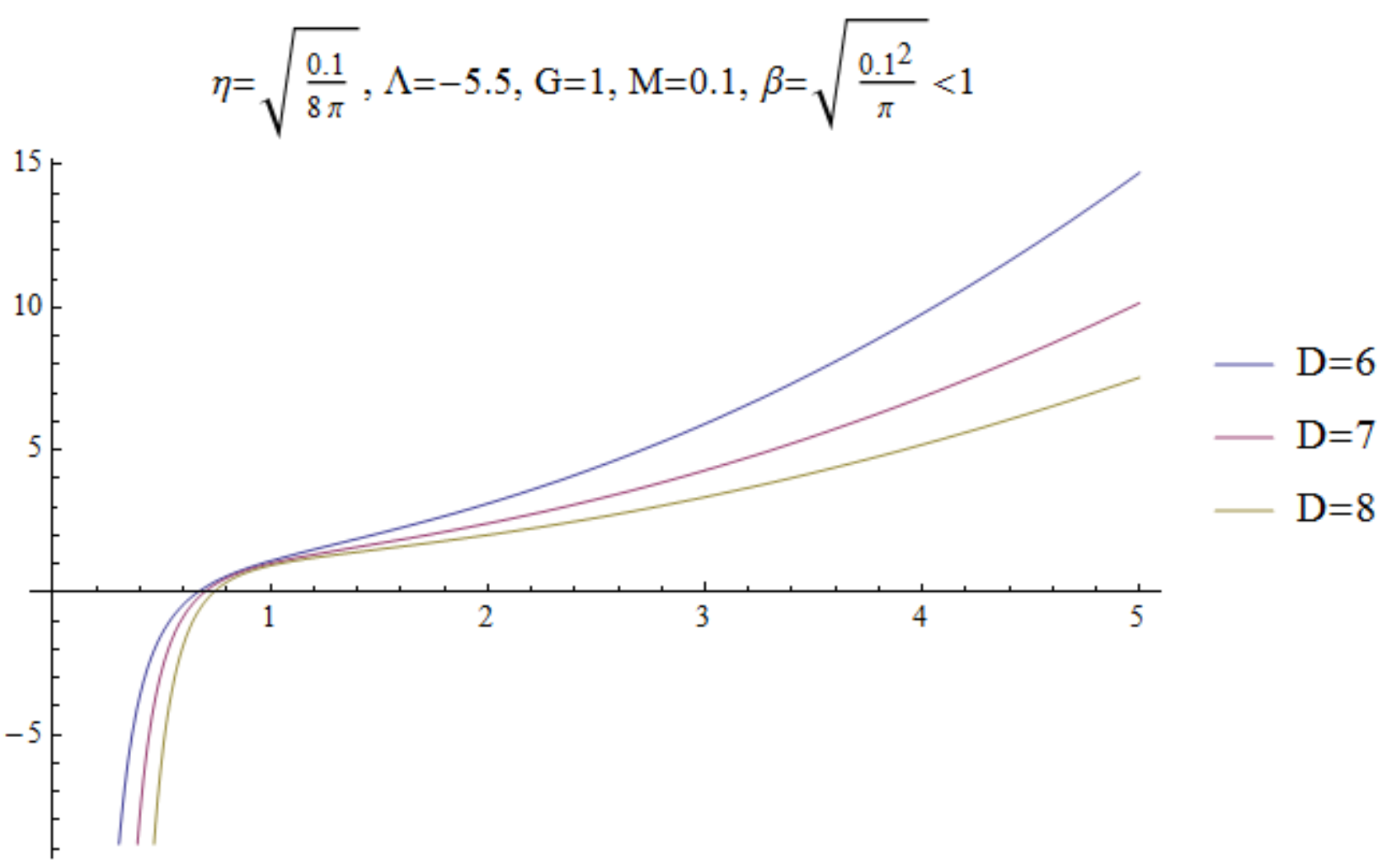}
	\caption{Existence of horizon in $D=6, 7, 8$ black holes with negative $\Lambda$.}
	\label{fig:k-ads-beda}
\end{figure}

\begin{figure}
	\centering
	\includegraphics[width=0.5\linewidth]{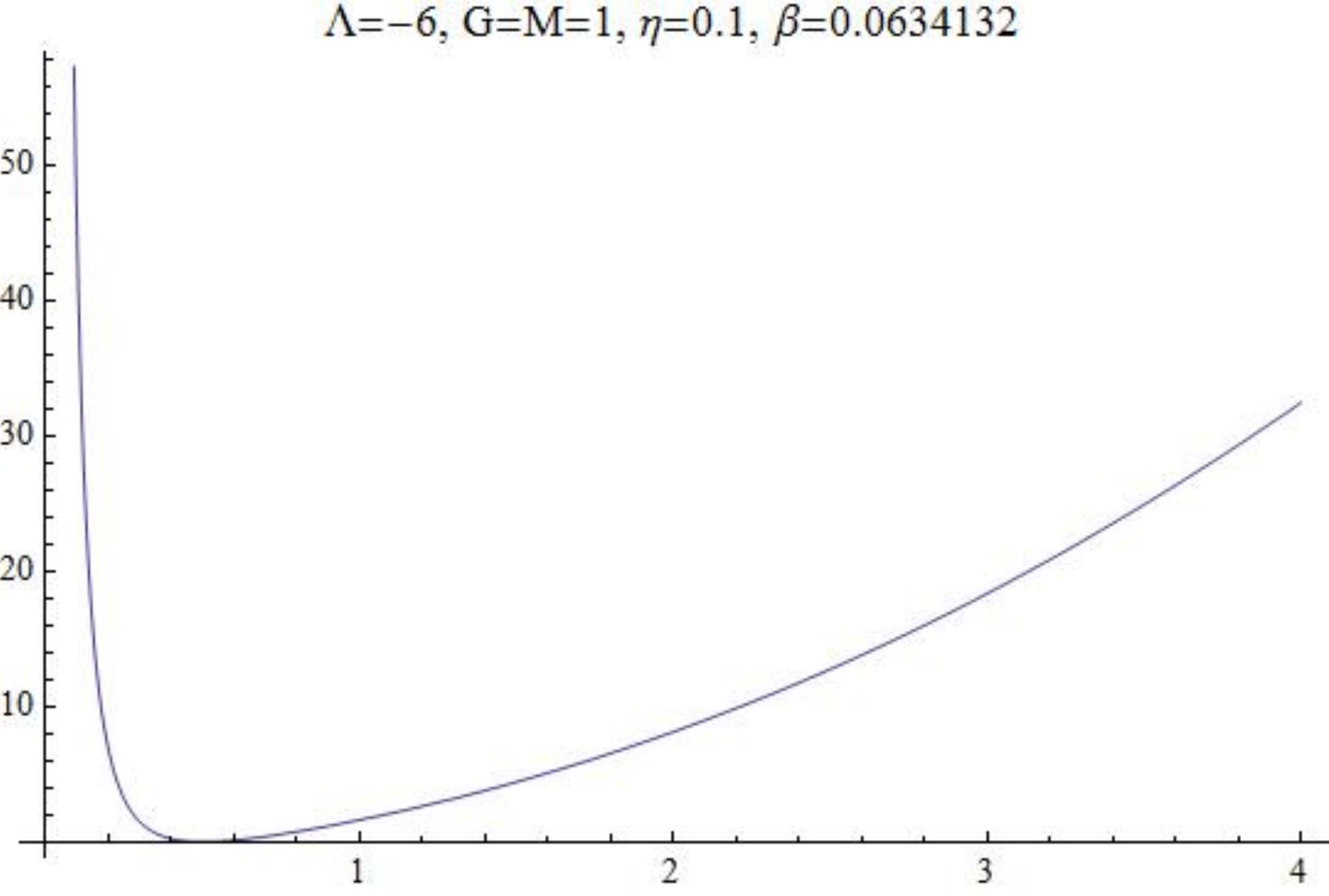}
	\caption{Extremal black hole in $D=4$ with negative $\Lambda$.}
	\label{fig:k-ads-ekstrem-d4}
\end{figure}
Notice that the second term is always positive. This results in the existence of only one horizon $r_+$ for $D>5$. They all reduce to the Tangherlini. This can be seen in Fig.~\ref{fig:k-ads-beda}. For $D=4$, however, we have a positive third term. This contribution allows the existence of two horizons, $r_-$ and $r_+$. The extremal condition can easily be deduced from Eq.~\eqref{eq:extremecold}. Re-writing it 
\begin{equation}
f(r)=\left(1-{r_0\over r}\right)^2\left[1+{|\Lambda|\over3}(r^2+2r_0 r+3r_0^2)\right],
\end{equation}
we can see that there exists at most one extremal condition, provided conditions~\eqref{eq:condextrem} holds with negative $\Lambda$,
\begin{eqnarray}
M&=&{r_0\over G}\left(1+{2\over3}|\Lambda| r_0^2\right),\nonumber\\
{\eta^4\over\beta^2}&=&{r_0^2\over8\pi G}\left(1+|\Lambda| r_0^2\right).
\end{eqnarray}
Such a solution is shown in Fig.~\ref{fig:k-ads-ekstrem-d4}.

Note that throughout this paper we only study black hole with spherical topology ($k=1$). It is well-known that AdS Reissner-Nordstrom black holes can exist with non-spherical topology (having $k=0,$ or $-1$), for example in~ \cite{Birmingham:1998nr, Chamblin:1999tk}. To the best of our knowledge the similar solutions having global monopole charge have yet been obtained. 

\section{Factorized solutions}

Another class of solutions to consider is when we set $C(r)=C=const$\footnote{This ansatz is similar to the case of, and can be regarded as solutions, spacetime compactification due to scalar~\cite{GellMann:1984sj}. The latter interpretation can be regarded as being generated by the nonlinearity of the scalar field's kinetic terms.}. The Einstein's equations become
\begin{eqnarray}
{(D-2)(D-3)\over2C^2}&=&\Lambda+{4(D-2)\pi G\eta^2\over C^2}+{2(D-2)^2 \pi G\eta^4\over C^4\beta^2},\label{eq:compacX1}\\
-{1\over B^2}{A''\over A}+{1\over B^2}{A'B'\over AB}+{(D-4)(D-3)\over2C^2}&=&\Lambda+{4(D-4)\pi G\eta^2\over C^2}+{2(D-6)(D-2)\pi G\eta^4\over C^4\beta^2}.\label{eq:compacX2}~~~~~~
\end{eqnarray}
From $G^0_0$, Eq.~\eqref{eq:compacX1}, we can solve for the radius
\begin{equation}
C^2={4(D-2)\pi G \eta^4 \over (D-3-8\pi G\eta^2)\beta^2}\label{eq:radiusnol}
\end{equation}
for $\Lambda=0$, and
\begin{equation}
C^2_\pm={(D-2)[(D-3-8\pi G\eta^2)\pm \sqrt{(D-3-8\pi G\eta^2)^2 -32\pi G \eta^4 \Lambda \beta^{-2}}]\over 4\Lambda}\label{eq:radiustaknol}
\end{equation}
for $\Lambda\neq 0.$ One can verify that as $\beta\rightarrow\infty$, $C_+^2\rightarrow{(D-2)(D-3-8\pi G\eta^2)/2\Lambda},$ approaching the OV solution~\cite{Olasagasti:2000gx}; when $\Lambda=0$ the compactification radius is not fixed by the theory. To ensure that $C$ is real it requires
\begin{equation}
\eta\leq{1\over\sqrt{8}}\sqrt{\frac{2 \beta ^2 (D-3)}{2 \pi  \beta ^2 G-\Lambda }- \sqrt{\frac{2\beta ^2 (D-3)^2 \Lambda }{\pi G \left(\Lambda -2 \pi  \beta ^2 G\right)^2}}}
\equiv\eta_{crit2}
\label{eq:etacrit2}
\end{equation}
for $\Lambda<2 \pi G\beta^2$,
\begin{equation}
\eta\leq{\eta_{crit}\over\sqrt{2}}\equiv\eta_{crit3},\label{eq:etacrit3}
\end{equation}
for $\Lambda=2 \pi G\beta^2$, and 
\begin{equation}
\eta \leq {1\over\sqrt{8}} \sqrt{\frac{2 \beta ^2 (D-3)}{2 \pi  \beta ^2 G-\Lambda }+\sqrt{\frac{2\beta ^2 (D-3)^2 \Lambda }{\pi G \left(\Lambda -2 \pi  \beta ^2 G\right)^2}}}
\equiv\eta_{crit4},
\label{eq:etacrit4}
\end{equation}
for $\Lambda>2\pi G\beta^2$. They satisfy
\begin{equation}
\eta_{crit3} < \eta_{crit2} < \eta_{crit4} < \eta_{crit}.
\end{equation}

Eq.~\eqref{eq:compacX2} can be re-written as
\begin{equation}
B^{-2}\left({A'B'\over AB}-{A''\over A}\right)=\omega^2, \label{pers03}
\end{equation}
where we define a constant $\omega$ 
\begin{eqnarray}
\omega^2&\equiv& \Lambda-{(D-4)(D-3-8\pi G\eta^2)\over 2C^2}+{2(D-6)(D-2)\pi G\eta^4\over C^4\beta^2}. %\nonumber\\
%&=& \Lambda-{\beta^2(D-4)(D-3-8\pi G\eta^2)^2\over 16\pi G\eta^4(D-2)}
%\left(1+\sqrt{1-{32\Lambda\pi G\eta^4\over \beta^2(D-3-8\pi G\eta^2)^2}}\right)\nonumber\\ 
%&&+{\beta^2(D-6)(D-3-8\pi G\eta^2)^2\over 32\pi G\eta^4(D-2)}\left[\left(1+\sqrt{1-{32\Lambda\pi G\eta^4\over \beta^2(D-3-8\pi G\eta^2)^2}}\right)\right]^2.
\end{eqnarray}
In general, $\omega^2$ can be positive, zero, or negative. Now, Eq.~(\ref{pers03}) can be solved, by taking an ansatz $B\equiv A^{-1}$, to give\footnote{Another choiec of ansatz, $B=1$, will lead to
\begin{eqnarray}
ds^2=
\begin{cases}
\frac{1}{\omega^2}(\sin^2 \chi ~dt^2 - d\chi^2) -C^2 d\Omega_{D-2}^2, & \text{for $\omega^2>0$},\nonumber\\
dt^2 -dr^2 -C^2 d\Omega_{D-2}^2, & \text{for } \omega=0,\nonumber\\
\frac{1}{\omega^2}(\sinh^2 \chi ~dt^2 - d\chi^2) -C^2 d\Omega_{D-2}^2, & \text{for $\omega^2<0$},\nonumber
\end{cases}\label{eq:metrsol1}
\end{eqnarray}
with $\chi\equiv\omega r$, which is essentially the same Nariai, Plebanski-Hacyan, or Bertotti-Robinson spacetimes as above, written in different gauge.} 
\begin{eqnarray}
ds^2=
\begin{cases}
(1 -\omega^2 r^2)~dt^2 -\frac{dr^2}{(1 -\omega^2 r^2)} - C^2 d\Omega_{D-2}^2, & \text{for $\omega^2>0$},\\
dt^2 -dr^2 - C^2 d\Omega_{D-2}^2, & \text{for } \omega=0,\\
(1 +\omega^2 r^2)~dt^2 -\frac{dr^2}{(1 +\omega^2 r^2)} - C^2 d\Omega_{D-2}^2, & \text{for $\omega^2<0$}.
\end{cases}\label{eq:metrsol2}
\end{eqnarray}
The resulting spacetimes are Naria ($dS_{2}\times S^{D-2}$)~\cite{Nariai}, Plebanski-Hacyan ($M_{2}\times S^{D-2}$)~\cite{Plebanski}, and Bertotti-Robinson ($AdS_{2}\times S^{D-2}$)~\cite{Bertotti:1959pf, Robinson:1959ev}, respectively. Here, $\omega^2$ plays the role of the effective cosmological constant in the two-dimensional maximally-symmetric  spacetimes. 
\begin{table}[h]
	\caption{Conditions for $k$-monopole compactification in $D$ dimensions.}
	\begin{tabular}{cccc}
		\hline \rule[-2ex]{0pt}{5.5ex}  & \,\,$dS_2\times S^{D-2}$\,\, & \,\,$M_2\times S^{D-2}$\,\, & \,\,$AdS_2\times S^{D-2}$ \\ 
		\hline \hline \rule[-2ex]{0pt}{5.5ex} $\Lambda>2 \pi  \beta^2  G$\ \ \; & $\eta^2 < \eta^2_{crit4}$ \ \  &\ \  $\eta^2 = \eta^2_{crit4}$ \ \ \ \ &\ \  cannot happen \ \  \ \ \\ 
		\hline \rule[-2ex]{0pt}{5.5ex} $\Lambda=2 \pi  \beta^2  G$\ \ \; & $\eta^2<\eta^2_{crit3}$ \ \  &\ \  $\eta^2=\eta^2_{crit3}$ \ \ \ \ &\ \  cannot happen \ \  \ \ \\ 
		\hline \rule[-2ex]{0pt}{5.5ex} $\Lambda<2 \pi  \beta^2  G$\ \ \; & $\eta^2 < \eta^2_{crit2}$ \ \  &\ \  $\eta^2 = \eta^2_{crit2}$ \ \ \ \ &\ \  cannot happen \ \  \ \ \\
		\hline \rule[-2ex]{0pt}{5.5ex} $\Lambda=0$\ \ \; & cannot happen & cannot happen \ \ \ \  &\ \  $\eta^2 < \eta^2_{crit}$  \ \ \ \ \\ 
		\hline \rule[-2ex]{0pt}{5.5ex} $\Lambda<0$\ \ \; & \; {} \; & \; cannot happen \ \ \ \  \; & \; {} \ \ \ \ \; \\ 
		\hline 
	\end{tabular} 
	\label{table:tangk}
\end{table}

To ensure which factorized channel can take place, we need to check whether the condition that satisfies $\omega^2$ simultaneously also holds for the positivity of $C^2$. This is done by solving the polynomial equations of $\omega^2>0$, $\omega^2=0$, or $\omega^2<0$, to obtain the allowed range of $\beta$ and $\eta$. Combining the results with the constraints of given by~\eqref{eq:etacrit}, \eqref{eq:etacrit2}, \eqref{eq:etacrit3}, and \eqref{eq:etacrit4}, we conclude that the (classically-)allowed conditions for compactifications are shown in Table~\ref{table:tangk}. Note that not all nine possibilities\footnote{$X_D\rightarrow Y_2\times S^{D-2}$, where $X$ and $Y$ can each stands for the de Sitter, Minkowski, or Anti-de Sitter.} can happen. Take, for example, the case of $AdS_2\times S^{D-2}$ compactification with $\Lambda<2 \pi  \beta^2  G$. Solving the polynomial $\omega^2<0$ yields $\eta^2 > \eta^2_{crit3}$. But this contradicts constraint~\eqref{eq:etacrit3}.  We therefore conclude that such compactification cannot happen.  
These imply the following possible channels.
\begin{eqnarray}
dS_D &\longrightarrow&
\begin{cases}
dS_2\times S^{D-2},\\
M_2\times S^{D-2};
\end{cases}\\
M_D &\longrightarrow&
AdS_2\times S^{D-2}.
\end{eqnarray}
%_________________________________
%\clearpage
\section{Conclusions}

We extend our previous investigation of $4d$ black hole and spacetime compactification solutions of $k$-global monopole, \cite{Prasetyo:2015bga}, into higher dimensions. This is done by studying the $D$-dimensional Einstein-$\sigma$-model theory with cosmological constant, where the scalar fields have noncanonical kinetic term; specifically in the form of power-law. In this work we assume the simplest ansatz, the spherically-symmetric hedgehog. The scalar field equation is then satisfied automatically, while the Einstein's equations take the form that describes the gravitational field outside the power-law global monopole. 

Our study reveals a rich spectrum of exact solutions, even in this simplest ansatz. The first type we obtain is a sclarly-charged black hole solution. This can be thought of as a noncanonical global monopole being eaten up by a blackhole. That there exist nontrivial blackhole solutions for this Einstein-scalar theory is not surprising since they evade the no-scalar-hair theorem due to the non-flatness of our asymptotic spacetime, even in the condition of $\Lambda=0$. As in the case of Barriola-Vilenkin (BV) or Olasagasti-Vilenkin  (OV) solutions, the spacetime around global monopole suffers from a deficit solid angle that grows with dimension, $\Delta\equiv 8\pi G\eta^2/(D-3)$. For our specific model, the solution is not valid for $D=5$ but behaves regularly otherwise. For $D=3$ it seems that the deficit angle causes the metric to blow up. But this is misleading, since this case should be treated separately. This is because for $D=3$ the internal manifold is $S^1$, which is flat, while $S^{D-2}$ (for $D>3$) is non-flat.

The genuine feature of our solution is the appearance of ``scalar" charge, which cannot be rescaled away by coordinate rescaling. In four dimensions, this charge resembles the Reissner-Nordstrom electromagnetic charge. With positive cosmological constant, there are three horizons and three corresponding extremal conditions; the cold, ultracold, and Nariai blackholes.  Our main results in this section is when $D>5$. In higher dimensions the charge term is negative with a fixed power of $r^{-2}$. Due to this peculiar property there are at most two horizons. We also found the higher-dimensional Nariai black hole solution with global power-law monopole within this spectrum, where the corresponding mass and scalar charge are given by~\eqref{eq:condextremD}. For $\Lambda<0$ all solutions, except at $D=4$, have only one horizon. For $4d$ case, however, the positivity of the scalar charge  enables the existence of two horizons, out of which we obtain one extremal black hole. It is well-known in the literature that a $(2+1)$-dimensional black hole can exist with $\Lambda<0$~\cite{Banados:1992wn, Banados:1992gq}. To the best of our knowledge the only obtained solutions for BTZ black holes having global monopole is by Mazharimousavi and Halilsoy~\cite{Mazharimousavi:2014uya}. The ``monopole" (or rather vortex) has $SO(2)$ global symmetry. In the next work we shall investigate such BTZ black holes in the context of noncanonical defects.

In this work we investigate the case for asymptotically anti de-Sitter black hole only in spherical topology ($k=1$). Higher-dimensional AdS black holes with noncanonical global monopole having planar ($k=0$) or hyperbolic ($k=-1$) topology is being studied at the moment. As mentioned in the text above, our investigation on hyperbolic monopole reveals the existence of surplus solid angle~\cite{ramadhanpradhana}. It is well-known that surplus angle is a generic property of Horava-Lifshitz gravity (\cite{Horava:2009uw, Horava:2008ih, Horava:2009if}) black holes (see, for example,~\cite{Kim:2009dq, Kim:2010vs, Kim:2010af}). Interestingly enough, the global monopole in Horava-Lifshitz gravity appears to have deficit solid angle~\cite{Lee:2010er}. It might be worth-investigating to consider this higher-dimensional noncanonical defects in the framework of Horava-Lifshitz gravity.    

Another type we obtain is factorized solutions, where the spacetime is the direct product of a two-dimensional Lorentzian manifold and an $(N-2)$-dimensional space, both of constant curvatures. For $\Lambda>0$ we can have Nariai ($dS_2\times S^{D-2}$) or Plebanski-Hacyan ($M_2\times S^{D-2}$) types, while for $\Lambda=0$ we only have Bertotti-Robinson ($AdS_2\times S^{D-2}$). There is no compactification in the case of $\Lambda<0$, though this conclusion might change should we work on planar or hyperbolic topologies. This possibilities is worth further investigation. We also found that all compactifications happen before the monopole reaches the super-critical stage. This is consistent to what we found previously for $D=4$ (see the erratum of~\cite{Prasetyo:2015bga}). We do not obtain any result when $\eta>\eta_{crit}$. This means that the fate of super-critical power-law global monopole cannot be probed analytically. It is interesting to study numerically the gravity of this higher-dimensional power-law global monopole. It might give bounds on $\eta$ of when should the compactification starts to develop, as in the case of canonical global monopole studied in~\cite{Liebling:1999bb}. 

There are several things still left undiscussed in this work. Even though we claim that the exact solutions we found are black hole with global defects, we have not investigated the stability condition. It may be that some, or even all, of them are perturbatively unstable. Studying the black hole's classical stability is a cumbersome task. We shall return to this issue in the subsequent publication. Another thing we have not said anything about is the thermodynamics properties of these solutions. We also only focus on a particular choice of $k$-monopole, a quadratic power-law. Despite the rich spectrum of solutions even in this simple theory, nothing should prevent us from considering other types of noncanonical monopole. In our previous work we studied another type of noncanonicality; that is in the form of  DBI kinetic term~\cite{Prasetyo:2016nng}. Its generalization to higher dimensions have also been worked and our investigation revealed several interesting properties that are genuine. We decided to report it in a separate publication~\cite{nextpub}. It is also interesting to see quantum tunneling in this simple landscape of vacua, as in~\cite{BlancoPillado:2009di, BlancoPillado:2009mi}. These issues, at the moment, are being worked on and we hopefully shall have any result to report soon.
%_______________________________________________
\section{Acknowledgements}

We thank Ardian Atmaja, Jose Blanco-Pillado, Haryanto Siahaan, Anto Sulaksono, and Alexander Vilenkin for useful comments and fruitful discussions. This work is partially supported by the ``PITTA" grant from Universitas Indonesia under contract no. 656/UN2.R3.1/HKP.05.00/2017.
%This work was partially supported by the University of Indonesia's Research Cluster Grant on ``Non-perturbative phenomena in nuclear astrophysics and cosmology'' No~1862/UN.R12/HKP.05.00/2015.

%===========================================================

\end{document}